\documentclass[aps,twocolumn,PRB,reprint,superscriptaddress]{revtex4}
\usepackage{amsmath}
\usepackage[percent]{overpic}
\usepackage{times}
\usepackage{subfigure}
\usepackage{latexsym}
\usepackage{graphicx}
\usepackage{bm}
\usepackage{amssymb}
\usepackage{anyfontsize}
\usepackage{hyperref}
\usepackage{multirow}
\usepackage{booktabs}
\usepackage{makecell}
\begin{document}
\title{Spin and charge transport in topological nodal-line semimetals}
\author{Yao Zhou}
\affiliation{National Laboratory of Solid State Microstructures, Department of Physics, Nanjing University, Nanjing 210093, China}
\author{Feng Xiong}
\affiliation{National Laboratory of Solid State Microstructures, Department of Physics, Nanjing University, Nanjing 210093, China}
\author{Weipeng Chen}
\affiliation{National Laboratory of Solid State Microstructures, Department of Physics, Nanjing University, Nanjing 210093, China}
\author{Jin An}
\email{anjin@nju.edu.cn}
\affiliation{National Laboratory of Solid State Microstructures, Department of Physics, Nanjing University, Nanjing 210093, China}
\affiliation{Collaborative Innovation Center of Advanced Microstructures, Nanjing University, Nanjing 210093, China}
\begin{abstract}
\end{abstract}

\date{\today}

\begin{abstract}
  We study transport properties of topological Weyl nodal-line semimetals(NLSs). Starting from a minimal lattice model with a single nodal loop, and by focusing on a normal-metal-NLS-normal-metal junction, we investigate the dependence of the novel transport behavior on the orientation of the nodal loop. When the loop is parallel to the junction interfaces, the transmitted current is found to be nearly fully spin-polarized. Correspondingly, there exists a spin orientation, along which the incident electrons would be totally reflected. An unusual resonance of half transmission with the participation of surface states also occurs for a pair of incident electrons with opposite spin orientations. All these phenomena have been shown to originate from the existence of a single forward-propagating mode in the NLS of the junction, and argued to survive in more generic multi-band Weyl NLSs.
\end{abstract}
	
\maketitle
\section{Introduction}
Topological Dirac or Weyl NLSs\cite{burkovTopological2011,phillipsTunable2014,xieNew2015,kimDirac2015,yuTopological2015,chenNanostructured2015,yamakageLineNode2015,mullenLine2015,fangTopological2015,chanMathrmCa2016,liDirac2016,zhaoTopological2016,chiuClassification2016,ezawaLoopNodal2016,bianDrumhead2016,bianTopological2016,schoopDirac2016,neupaneObservation2016,huEvidence2016,takaneDiracnode2016,wuDirac2016,quanSingle2017,sunDirac2017,limPseudospin2017,zhangTransport2017,huQuantum2017,yangSymmetry2018,kimSpinorbitstable2018,liRules2018,armitageWeyl2018,ahnBand2018,takaneObservation2018,louExperimental2018,liuExperimental2018,liEvidence2018,fuDirac2019,qiuObservation2019,lauTopological2019,bhowalDirac2019,guoElectronic2019}, characterized by one-dimensional(1D) band crossings between bulk conduction and valence bands in momentum space, and topologically protected drumhead surface states\cite{burkovTopological2011,armitageWeyl2018} at the boundary, have attracted much attention recently. Although some materials are proposed to exhibit nodal-line fermions near the Fermi level, only a few of them including XTaSe$_{2}$(X=Pb, Tl)\cite{bianDrumhead2016,bianTopological2016}, ZrXY(X=Si,Ge; Y=S, Se, Te)\cite{schoopDirac2016,neupaneObservation2016,huEvidence2016,zhangTransport2017,huQuantum2017,liRules2018,fuDirac2019,guoElectronic2019}, PtSn$_{4}$\cite{wuDirac2016}, XB$_{2}$(X=Al, Zr, Ti)\cite{takaneObservation2018,louExperimental2018,liuExperimental2018}, SrAs$_{3}$\cite{liEvidence2018} have been experimentally verified by angle-resolved photoemission spectroscopy(ARPES) and quantum oscillations. NLS materials always demonstrate rich topological configurations such as nodal-net\cite{yuTopological2015,wengTopological2015,shaoTunable2018,liuExperimental2018,fengTopological2018}, nodal-chain\cite{bzdusekNodalchain2016,zhangIdeal2018,caiNodalchain2018}, and Hopf-link\cite{chenTopological2017,yanNodallink2017,ezawaTopological2017,changWeyllink2017,biNodalknot2017,ahnBand2018,zhouHopflink2018} structures formed by the nodal lines. Among these materials, most are Dirac NLSs, in which the line nodes are four-fold degenerate. PbTaSe$_{2}$\cite{bianTopological2016} is an exception, which is a spin-orbit(SO) Weyl NLS possessing several doubly degenerate nodal lines. Very recently, the room-temperature magnet Co$_{2}$MnGa\cite{belopolskiDiscovery2019} with negligible SO interaction has been discovered to be a Weyl NLS, exhibiting exotic transport effects.

Up to now, a majority of the transport experiments\cite{huEvidence2016,huQuantum2017,zhangTransport2017,liRules2018,liEvidence2018,guoElectronic2019} have mainly focused on the confirmation of the existence of the nodal lines in the bulk materials, less concern has been shown to the novel nature itself of the transport property in the NLS materials. Theoretically, spin related transport properties have been studied for the NLSs. Phenomena such as resonant spin-flipped reflection\cite{chenProposal2018} and anomalously Hall current\cite{ruiTopological2018,martin-ruizParity2018} were predicted. In this paper, we study the spin and charge transport in the Weyl NLSs with a single nodal loop. For a junction made up of the NLS and normal metals, we find that due to the existence of only one forward-propagating mode in the Weyl NLS region, exotic transport phenomena occur. The transmitted charge current is found to be nearly fully spin-polarized. For a relevant scattering state, total reflection and surface states involved transmission resonance are found, the latter of which is also accompanied by the half transmission. These unusual features are expected to be verified in the future transport experiments in Weyl NLS materials.

This paper is organized as follows. In Sec. \ref{sect2}, based on a lattice model, we introduce a new wave-function transport method and then derive the related conservation laws as well as the corresponding charge and spin current densities and spin torque. In Sec. \ref{sect3}, starting from a minimal model of NLS, we successively discuss the transport properties of the N-NLS-N(where N represents normal metal) junctions when the nodal loop is parallel or perpendicular to the interfaces, or intersecting them at $45^{\circ}$. The nonconserved spin current density and spin torque together with the extension to multi-band case are also discussed. In Sec. \ref{sect4}, we summarize our results.

\section{Transport Method For a Lattice System}
\label{sect2}
\subsection{Scattering matrix}
\begin{figure}[ht]
  \begin{center}
	\includegraphics[width=7cm,height=7cm]{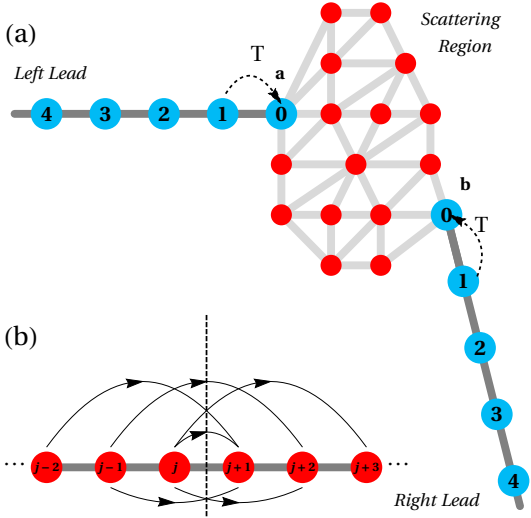}	
  \end{center}
  \vspace{-0.4cm}
  \caption{(Color online) (a)Schematic of a two-terminal lattice transport system, where the centered scattering region is so constructed that it contains the end unit cells $a$ and $b$ of the half-infinite normal leads. (b)A quasi-1D lattice, which is part of the transport system. The dashed line denotes a cross section between neighboring site $j$ and $j+1$, while the six curved arrows represent the bond currents whose paths intersect with the dashed line. Here the bonds up to the $3$rd NN hoppings are taken into account.
  } \label{fig1}
\end{figure}

In this section, we introduce a wave-function method in lattice form, similar to Ref.\cite{grothKwant2014}, to solve the transport problem of a noninteracting scattering system with multiple terminals. In the following we take a two-terminal case as an example, schematically shown in Fig.\ref{fig1}, to illustrate the essential points of the method. Both normal leads are assumed to be translational invariant along the propagating directions, and can be viewed as quasi-one-dimensional(1D) half-infinite lattices. Given an energy $E$, one can obtain for each lead all modes $\phi_{m}$, characterized by their wave vectors $\pm k_{m}$, $m=1,2,...M'$, with 2$M'$ being the total number of the modes. For simplicity, we assume that $k_{m}$ is real when $m\leq M$, corresponding to the propagating modes, whereas $k_{m}$ is complex when $M<m\leq M'$, corresponding to evanescent ones. The wave functions are the superpositions of all possible modes. For the scattering state of the propagating mode $\phi_{n}(n\leq M)$ incoming from lead $L$, the spinor wave functions at site $j$($j\geq0$)of lead L and R can be given by,
\begin{align}
  \Psi_{n}^{L}(j) &= \phi_{n} e^{-i k_{n} j}  + \sum^{M'}_{m=1} r_{mn} \phi_{m} e^{i k_{m} j}, \label{leadfunc1} \\
  \Psi_{n}^{R}(j) &=  \sum^{M'}_{m=1} t_{mn} \phi_{m} e^{i k_{m} j},
  \label{leadfunc2}
\end{align}
where $r_{mn}$ and $t_{mn}$ are the reflection and transmission amplitudes. Here for each propagating mode $m$, $\phi_{m}$ is so normalized that its group velocity $v_{m}=\langle\phi_{m}\mid\partial_{k}H_{L(R)}(k)\mid_{k=k_{m}}\mid\phi_{m}\rangle$ is fixed to be $1$, where $H_{L(R)}(k)$ is the Hamiltonian of lead $L(R)$.

On the other hand, the wave function $\Psi$ of the whole transport system should obey the stationary Schr\"odinger equation
\begin{equation}
  \mathcal{H} \Psi = E \Psi,
  \label{Hequ}
\end{equation}
where $\Psi=(\Psi^{S}, \dots, \Psi^{L}(j), \dots, \Psi^{R}(k), \dots)^{T}$, and $\mathcal{H}$ takes the following form,
\begin{equation}
  \mathcal{H} =
  \left[
	\begin{array}{c|c|c}
	  \mbox{\Huge{$H_{S}$}} &  \begin{array}{ccc}
							T & \cdots  & 0  \\
							0 & \vdots & \vdots \\
							0 & 0   & 0  \\
						  \end{array} & \begin{array}{ccc}
                                          T & \cdots  & 0  \\
                                          0 & \vdots & \vdots \\
                                          0 & 0   & 0  \\
                                        \end{array} \\             \hline
	  \begin{array}{ccc}
		T^{\dagger} & 0  & 0  \\
		\vdots  & \vdots & 0 \\
		0 & \cdots   & 0  \\
		\end{array} & \mbox{\Huge{$H_{L}$}}  & \mbox{\Huge{0}}

												 \\    \hline
	  \begin{array}{ccc}
	  T^{\dagger} & 0  & 0   \\
	   \vdots  & \vdots & 0 \\    					
	   0 & \cdots   & 0  \\
	   \end{array}     &  \mbox{\Huge{0}}   & \mbox{\Huge{$H_{R}$}} \\
	\end{array}
  \right].
\end{equation}
Here $H_{S}$ and $\Psi^{S}$ are the Hamiltonian matrix and wave function of the scattering region. $T$ is the nearest-neighboring(NN) hopping matrix in the normal leads. The Schr\"odinger Eq.(\ref{Hequ}) is composed of three sets of equations, among which the last two are obeyed exactly by $\Psi^{L}$ and $\Psi^{R}$ described in Eq.(\ref{leadfunc1})-(\ref{leadfunc2}). The remaining unknown $\Psi^{S}$ can be solved as follows,
\begin{equation}
  \Psi^{S} = G(E)
  \underbrace{
  \begin{bmatrix}
	T\Psi^{L}(1)\\
	0\\
	\vdots  \\
	0\\
	T\Psi^{R}(1) \\
  \end{bmatrix}
}_{N \times 1},
\end{equation}
where $G(E)=(E-H_{S})^{-1}$ is the Green's function of the scattering region. More explicitly, $\Psi^{S}$ can be written as,
\begin{equation}
  \Psi^{S}(j)=G_{ja}(E)T\Psi^{L}(1) +G_{jb}(E)T\Psi^{R}(1).
  \label{scatfunc}
\end{equation}

As shown in Fig.\ref{fig1}, cell $a$($b$) in the scattering region is actually the $0$th cell of lead $L(R)$. Correspondingly, their wave functions should be identical to each other,
\begin{eqnarray}
  \Psi^{L}(0) &=& \Psi^{S}(a), \\
  \Psi^{R}(0) &=& \Psi^{S}(b).
\end{eqnarray}
According to the above equations, one can determine the coefficients $r_{mn}$ and $t_{mn}$, among which those associated with the propagating modes constitute a 2$M\times$2$M$ scattering matrix
\begin{equation}
  S =
  \begin{bmatrix}
	r & t^{'} \\
	t & r^{'}
  \end{bmatrix},
\end{equation}
where the $r(r^{'}), t(t^{'})$ are the $M\times M$ reflection and transmission matrix for the scattering state incident from lead L(R). Owing to current conservation, $S$ is unitary $S^{\dagger}S=SS^{\dagger}=1$, i.e.,
\begin{eqnarray}
  r^{\dagger}r + t^{\dagger}t=r'^{\dagger}r' + t'^{\dagger}t'&=I, \\
  rr^{\dagger} + t't'^{\dagger}=r'r'^{\dagger} + tt^{\dagger}&=I, \\
  r^{\dagger}t'+ t^{\dagger}r'=rt^{\dagger} + t'r'^{\dagger}&=0.
\end{eqnarray}
These conservation equations actually guarantee that in a time-reversal invariant system, transport property is Fermi surface's property. This means that for any scattering state with $E<E_{F}$, in any normal lead, all charge or spin current contributions from all possible modes with the same $E$(which may be incident from the same or different lead) would cancel each other out.

Compared with the Green-function transport method\cite{datta1997electronic}, this one is much more simple and gives directly the wave function of the scattering region, which is essential in calculating the transport physical quantities such as the spin and charge current densities.
\subsection{Conservation laws, charge and spin current densities}

Now we discuss the spin and charge conservation laws of a generic quasi-1D lattice system. Consider a tight-binding model with multiple NN hoppings, described by,
\begin{eqnarray}
  i\hbar\partial_{t} \Psi(j) = \epsilon_{j}\Psi(j)+\sum_{\delta=\pm1,\pm2,\cdots}T_{j,j+\delta} \Psi(j+\delta),
\end{eqnarray}
where the spinor wave function  $\Psi(j)$ contains both spin and orbital degrees of freedom. $T_{i,j}$ is the hopping matrix between site $i$ and $j$. $\epsilon_{j}$ is the on-site energy matrix, in which SO interactions might be included. Introducing the charge and spin densities $\rho(j)=e\Psi^{\dagger}(j)\Psi(j), \bm{S}(j)=\frac{\hbar}{2}\Psi^{\dagger}(j)\bm{\sigma}\Psi(j)$, together with the bond charge and spin current densities defined respectively by,
\begin{eqnarray}
J^{c}_{i,j}&=& \frac{2e}{\hbar} \operatorname{Im}\{\Psi^{\dagger}(i)T_{i,j}\Psi(j)\}, \\
\bm{J}^{s}_{i,j}&=& \operatorname{Im}\{\Psi^{\dagger}(i)\frac{1}{2}\{\bm{\sigma}, T_{i,j}\}\Psi(j)\}=J^{s}_{i,j}\bm{\hat{n}},
\end{eqnarray}
the continuity equations\cite{sunDefinition2005,soninEquilibrium2007,ruckriegelSpin2017} of our lattice model can be derived as,
\begin{eqnarray}
  \partial_{t}\rho(j) + J^{c}_{j+1\leftarrow j} - J^{c}_{j\leftarrow j-1}  &=&0, \label{continuity} \\
  \partial_{t}\bm{S}(j)+ \bm{J}^{s}_{j+1\leftarrow j} -\bm{J}^{s}_{j\leftarrow j-1} &=& \bm{g}(j).
  \label{continuity1}
\end{eqnarray}
Here $\bm{g}(j)$ is the spin torque term\cite{mishchenkoSpin2004,sunDefinition2005,soninEquilibrium2007,ruckriegelSpin2017,shiProper2006,ralphSpin2008,hattoriSpinCurrentDriven2009,brataasCurrentinduced2012,mahfouziSpinOrbit2012,wangSpinorbitcoupled2014,mellnikSpintransfer2014,nomuraChargeInduced2015,dyrdalCurrentinduced2015,zeleznySpin2018,dcRoomtemperature2018,fallahiSpin2019}, which plays an important role in SO coupled systems, and $J^{c}_{j+1\leftarrow j}(\bm{J}^{s}_{j+1\leftarrow j})$ is the charge(spin) current density flowing through the cross section between neighboring site $j$ and $j+1$. Either of the current densities can be expressed as the sum over the bond currents whose hopping paths intersect with the cross section. As an illustration, for a model system with up to the $3$rd NN hoppings, $J^{c}_{j+1\leftarrow j}$ can be expressed as,
\begin{equation}
\begin{split}
J^{c}_{j+1\leftarrow j}&=J^{c}_{j+3,j}+J^{c}_{j+2,j}+J^{c}_{j+2,j-1} \\
&+J^{c}_{j+1,j}+J^{c}_{j+1,j-1}+J^{c}_{j+1,j-2},
\end{split}
\end{equation}
which is schematically shown in Fig.\ref{fig1}(b). $\bm{J}^{s}_{j+1\leftarrow j}$ has the exactly similar expression. Meanwhile, the spin source term $\bm{g}(j)$ takes the following form,
\begin{equation}
  \begin{split}
	\bm{g}(j) &= \operatorname{Im}\{\Psi^{\dagger}(j)\frac{1}{2} \left[\bm{\sigma}, \epsilon_{j} \right] \Psi(j) ) \} \\
    &+ \sum_{\delta}\operatorname{Im}\{\Psi^{\dagger}(j)\frac{1}{2}\left[\bm{\sigma}, T_{j,j+\delta} \right]\Psi(j+\delta) \}.	
  \end{split}
  \label{source}
\end{equation}
If the transport system has SO coupling, $\bm{g}(j)$ is generally nonzero, so the spin current is not conserved. Furthermore, if the system is in a steady state, i.e., $\partial_{t}\rho(j)=\partial_{t}\bm{S}(j)=0$, a conserved charge current density $J^{c}\equiv J^{c}_{j+1\leftarrow j}=J^{c}_{j\leftarrow j-1}$ independent of the cross section's location can be defined, while the spin current $\bm{J}^{s}(j)\equiv \bm{J}^{s}_{j+1\leftarrow j}$ still depends on $j$ since it is generally not conserved. However, for the normal leads, which are assumed to be SO decoupled, $\bm{J}^{s}$ is also conserved. Therefore, for a scattering state $n$ described by Eq.(\ref{leadfunc1})-(\ref{leadfunc2}), its contribution to the charge and spin current density at both leads can be given as,

\begin{equation}
\begin{split}
J^{c}_{n,R}&=\frac{e}{\hbar}\sum^{M}_{m=1}\mid t_{mn}\mid^{2}=\frac{e}{\hbar}(t^{\dagger}t)_{nn}, \\
J^{c}_{n,L}&=\frac{e}{\hbar}\{1-\sum^{M}_{m=1}\mid r_{mn}\mid^{2}\}=\frac{e}{\hbar}\{1-(r^{\dagger}r)_{nn}\},
\end{split}
\end{equation}
\begin{equation}
\begin{split}
\bm{J}^{s}_{n,R}&=\sum^{M}_{m=1}\langle\bm{\sigma}\rangle_{m}\mid t_{mn}\mid^{2}=(t^{\dagger}\bm{\sigma}t)_{nn}, \\
\bm{J}^{s}_{n,L}&=\langle\bm{\sigma}\rangle_{n}-\sum^{M}_{m=1}\langle\bm{\sigma}\rangle_{m}\mid r_{mn}\mid^{2} \\
&=\langle\bm{\sigma}\rangle_{n}-(r^{\dagger}\bm{\sigma}r)_{nn},
\end{split}
\end{equation}
where $\langle\bm{\sigma}\rangle_{m}=\langle\phi_{m}\mid\bm{\sigma}\mid\phi_{m}\rangle/\langle\phi_{m}\mid\phi_{m}\rangle$. Generically, one has $J^{c}_{n,R}=J^{c}_{n,L}$ according to (\ref{continuity}), but $\bm{J}^{s}_{n,R}\neq\bm{J}^{s}_{n,L}$, due to the spin-torque effect in the NLS. The zero-biased charge(spin) conductance $G^{c}\equiv dI^{c}/dV(\bm{G}^{s}\equiv d\bm{I}^{s}/dV)$ is proportional to the sum of the current density contributed from each mode,
\begin{equation}
\begin{split}
G^{c}_{R}=\frac{e^{2}}{h}\mathrm{Tr}(t^{\dagger}t)=G^{c}_{L}=\frac{e^{2}}{h}\{M-\mathrm{Tr}(r^{\dagger}r)\},
\end{split}
\label{chargecond}
\end{equation}
\begin{equation}
\begin{split}
\bm{G}^{s}_{R}=\frac{e}{2\pi}\mathrm{Tr}(t^{\dagger}\bm{\sigma}t), \bm{G}^{s}_{L}=-\frac{e}{2\pi}\mathrm{Tr}(r^{\dagger}\bm{\sigma}r).
\end{split}
\end{equation}
Similarly, $\bm{G}^{s}_{R}\neq \bm{G}^{s}_{L}$.

\section{Transport properties in the N-NLS-N junction}
\label{sect3}
In general, the SO coupled systems often exhibit rich transport phenomena such as spin Hall effect\cite{murakamiDissipationless2003,sinovaUniversalIntrinsicSpin2004,sinovaSpinHallEffects2015} and spin-polarized current\cite{mishchenkoSpin2004,mahfouziSpinOrbit2012,dyrdalCurrentinduced2015}. On the other hand, the transport properties of a scattering system strongly depend on its Fermi-surface topology. The topological NLSs are thus expected to exhibit exotic features, since they usually have both nontrivial Fermi surface and strong SO coupling. In real NLS materials, there are always several nodal loops which could be linked or connected. These topologically nontrivial geometric structures may have great influence on the transport properties of the NLSs. In this paper, however, we are only focused on the NLSs with a single nodal loop. To eliminate other possible effect, we consider a minimal lattice model as follows,
\begin{figure}[ht]
  \begin{center}
	\includegraphics[width=7cm,height=4.7cm]{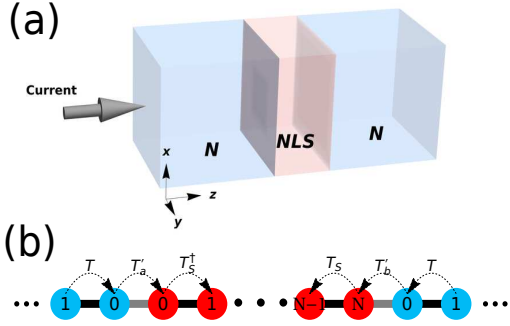}	
  \end{center}
  \vspace{-0.4cm}
  \caption{(Color online) (a)Schematic of the N-NLS-N junction, where two normal metals are connected to the centered NLS and charge current is flowing along $\bm{z}$ direction. (b)The effective quasi-1D lattice of the junction, when the translational invariances of the above system along $\bm{x}$ and $\bm{y}$ directions are taken in account. Here the blue and red circles represent the unit cells of the normal metals and the NLS respectively.
  } \label{fig2}
\end{figure}
\begin{equation}
  \mathcal{H}_{NLS}(\bm{k}) = (m-\cos k_{x}-\cos k_{y}-\cos k_{z} )\sigma_{x}+\lambda \sin k_{z}\sigma_{z},
  \label{paraNLS}
\end{equation}
where $m$, $\lambda$ are adjustable parameters. When $m$ is properly chosen to be within $1<m<3$, this minimal model describes a topologically nontrivial semimetal with one single nodal loop located at $k_{z}=0$ and $\cos k_{x}+\cos k_{y}=m-1$, which is independent of $\lambda$. This NLS is characterized by the drumhead surface states, which lie inside the projection of the nodal loop on the boundary and are protected by the chiral symmetry( $\{\sigma_{y},\mathcal{H}_{NLS}(\bm{k})\}=0$), $\mathcal{PT}$ symmetry( $\mathcal{K}\mathcal{H}_{NLS}(\bm{k})\mathcal{K}=\mathcal{H}_{NLS}(\bm{k})$ with $\mathcal{K}$ being the complex conjugation), and mirror symmetry( $\sigma_{x}\mathcal{H}_{NLS}(k_{x},k_{y},-k_{z})\sigma_{x}=\mathcal{H}_{NLS}(k_{x},k_{y},k_{z})$) held by the model system. In the following, $m=2.5$ is chosen to fix the radius $k_{L}$ of the circle-like loop to be about $\frac{\pi}{3}$.
\begin{figure}[ht]
  \begin{center}
	\includegraphics[width=8.5cm,height=7.5cm]{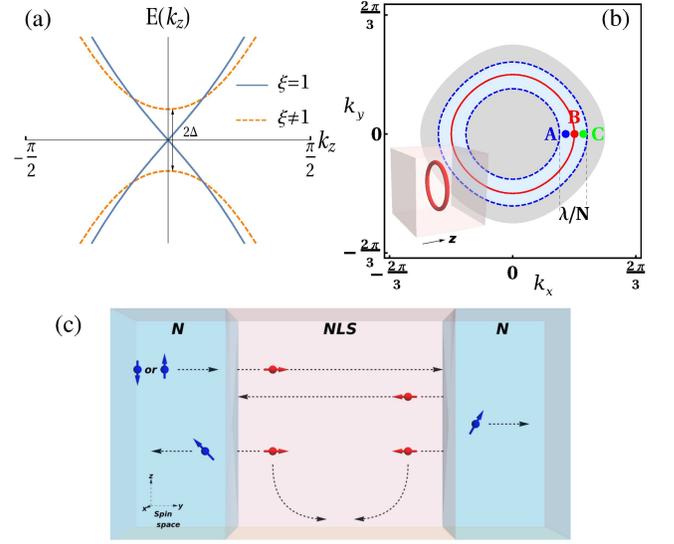}	
  \end{center}
  \vspace{-0.4cm}
  \caption{(Color online) (a)Low-energy dispersion of the effective 1D NLS in the parallel case for $\xi=1$ and $\xi\neq1$. While the former is gapless, the latter has a small gap $\Delta=|1-\xi|$. (b)The projection of the nodal loop on $k_{z}=0$ plane, represented by the solid loop, which together with the two dotted loops form a blue annulus. Electrons incoming from within(outside) this area give significant(vanishingly small) contributions to the transport. The width of the annulus is estimated to about $\lambda/N$, with $N$ being the number of layers of the NLS. $A$,$B$,$C$ are three representative incident points, with $\bm{k}^{\perp}_{B}=(\frac{\pi}{3},0)$, $\bm{k}^{\perp}_{A(C)}=(\frac{\pi}{3}+\delta k_{A(C)},0)$, and $\delta k_{A}=-0.01$, $\delta k_{C}=0.006$. The grey area represents the projection of the normal leads' Fermi sphere. The inset gives the 3D geometry of the nodal loop( $\perp \bm{z}$). (c)Schematic of the scattering mechanism for the fully spin polarization of the transmitted wave. The horizontal(curved) black arrows represent propagating(evanescent) modes, while the colored arrows represent spin orientations, whose coordinate frame is given by the inset in the lower left.
  } \label{fig3}
\end{figure}

To extract the transport features of this NLS, we turn to investigate its junction with normal metals. Below we concentrate mainly on the N-NLS-N junction, as schematically shown in Fig.\ref{fig2}(a). From the viewpoint of Sec.\ref{sect2}, the two normal metals can be seen as the normal leads, whose Hamiltonian is assumed to be
\begin{equation}
  \mathcal{H}_{N}(\bm{k}^{N}) = -2(\cos k^{N}_{x}+\cos k^{N}_{y}+\cos k^{N}_{z})-\mu,
\end{equation}
where $\mu$ is the chemical potential. In most of our calculations, we choose $\mu=-4$ to fix the radius of Fermi sphere $k^{N}_{F}$ to be about $\frac{\pi}{2}$, satisfying $k_{F}>k_{L}$. Since the whole transport system has translational invariances along both $\bm{x}$ and $\bm{y}$ directions, the N-NLS-N junction can be regarded as a quasi-1D one as demonstrated in Fig.\ref{fig2}(b). The total number of modes in this situation are 4: spin-up and spin-down forward-propagating ones and 2 corresponding backward-propagating ones, i.e., mode index $m=\uparrow$, $\downarrow$, ${M}'=M=2$ and $r$, $t$, ${r}'$, ${t}'$ become $2\times2$ matrices. From now on we start from this effective 1D scattering system to analyze the transport phenomena of the NLS. The hopping matrices $T'_{a}$, $T'_{b}$ between the normal leads and the NLS are also assumed to be $T'_{a}=T'_{b}=\gamma1_{2\times2}$ and in most of the calculations below we choose $\gamma=-1$.

\subsection{Nodal loop parallel to the interfaces}
We first consider the case where the nodal loop lies within $k_{z}=0$ plane, i.e., it's parallel to the interfaces. Since $k^{N}_{x}=k_{x}$ and $k^{N}_{y}=k_{y}$ are good quantum numbers, the effective 1D system is actually equivalent to Kitaev model\cite{kitaevUnpaired2001} describing a p-wave superconductor. The Hamiltonian and its energy dispersion can be given by,
\begin{align}
  \mathcal{H}^{\parallel}_{\mathrm{eff}}(k_{z}) &= (\xi - \cos k_{z} )\sigma_{x} + \lambda \sin k_{z}\sigma_{z}, \label{paraEH} \\
  E^{\parallel}(k_{z}) &= \pm \sqrt{(\xi - \cos k_{z})^2 + \lambda^{2} \sin^{2} k_{z}}, \label{paradisp}
\end{align}
where $\xi\equiv\xi(\bm{k}^{\perp})=m-\cos k_{x} -\cos k_{y}$ is a varying parameter depending on $\bm{k}^{\perp}\equiv(k_{x},k_{y})$ of the incident electrons. The 1D system $\mathcal{H}^{\parallel}_{\mathrm{eff}}(k_{z})$ is topologically nontrivial if $|\xi|<1$ and trivial otherwise. It is also characterized by the fact that it is gapless if $\xi=1$, but has a finite gap $\Delta=|1-\xi|$ otherwise, shown explicitly in Fig.\ref{fig3}(a). The incident electrons with relatively large $|1-\xi|$ are thus expected to have little contributions to the electron transport. Since $\xi=1$ corresponds to the projection loop, only the incoming electrons corresponding to the neighboring area of the loop($\xi-1\sim0$) contribute significantly to the transport. This area is exhibited as the blue annulus in Fig.\ref{fig3}(b).

\begin{figure}[ht]
  \begin{center}
	\includegraphics[width=8.5cm,height=9cm]{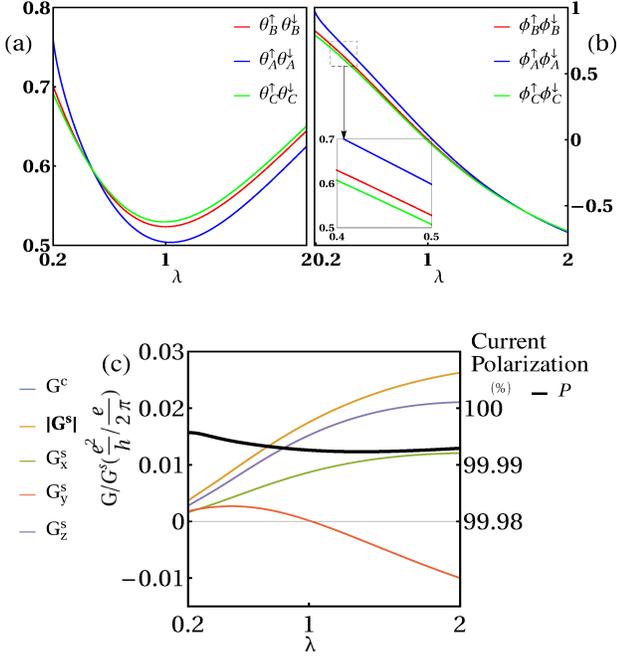}
  \end{center}
  \vspace{-0.4cm}
  \caption{(Color online) (a)-(b)Spin-polarization angle $(\theta,\phi)$ of the transmitted wave as a function of parameter $\lambda$. Data are shown for the three representative incident points denoted in Fig.\ref{fig3}(b). The spin polarization is found to be independent of incident spin orientation, so $\theta^{\uparrow}=\theta^{\downarrow},\phi^{\uparrow}=\phi^{\downarrow})$. (c)Transmitted charge and spin conductances, as well as current polarization as functions of parameter $\lambda$.
  } \label{fig4}
\end{figure}
Without loss of generality, consider $\lambda>0$. As demonstrated in detail in Appendix A, when $\xi-1\sim0$, $\mathcal{H}^{\parallel}_{\mathrm{eff}}(k_{z})$ has four eigensolutions: $k_{z}=\pm i\chi_{1}$, $\pm i\chi_{2}$, with $\chi_{1}=\frac{1-\xi}{\lambda}$ and $\chi_{2}=\ln(\frac{1+\lambda}{1-\lambda})$. The generic wave function can be written as,
\begin{equation}
  \begin{split}
	\Psi^{S}(j)&= (a e^{- \chi_{1}j} + b e^{-\chi_{2}j})\phi_{+}+(c e^{\chi_{1}j } + d e^{\chi_{2}j})\phi_{-},
  \end{split}
  \label{parawave}
\end{equation}
where $a,b,c,d$ are superposition coefficients, and $\phi_{\pm}=\frac{1}{\sqrt{2}}(1,\pm i)^{T}$. While $\chi_{2}$ is nearly independent of $\xi$ and corresponding to the evanescent states localized at the interfaces with their attenuation length $\chi^{-1}_{2}$ being approximately a few lattice constants, $\chi_{1}$ strongly depends on $\xi$ and even $\chi_{1}=0$ when $\xi=1$. For any finite length $N$ of the NLS system, as long as $\mid1-\xi\mid<\lambda/N$, the two evanescent states corresponding to $\chi_{1}$ can be actually viewed as propagating ones since their attenuation length $\chi^{-1}_{1}$ can be comparable with $N$.

\subsubsection{Fully spin-polarized transmitted current}
Consider the left-incoming scattering states with their $\bm{k}^{\perp}$ lying inside the annulus. For a spin-$\sigma$ incident electron, the transmitted wave can be written as
\begin{equation}
  \Psi^{R}(j) =
  \begin{bmatrix}
	t_{\uparrow\sigma} \\
	t_{\downarrow\sigma}
  \end{bmatrix}
  e^{-i kj}\equiv
  t^{\sigma}
  \begin{bmatrix}
	\cos\frac{\theta}{2}e^{-i\phi} \\
	\sin\frac{\theta}{2}
  \end{bmatrix}
  e^{-i kj}
  \label{paraRleadwave}
\end{equation}
Here the total transmission amplitude $t^{\sigma}$ of the transmitted wave is introduced with $(\theta,\phi)$ being its spin orientation, and $k=k_{\uparrow}=k_{\downarrow}\equiv k^{\mathrm{N}}_{z}=\cos^{-1}(\xi-0.5)\sim\frac{\pi}{3}$.

We now show that $(\theta,\phi)$ is independent of the incident electron's spin orientation $\sigma$, i.e., $(\theta^{\uparrow},\phi^{\uparrow})=(\theta^{\downarrow},\phi^{\downarrow})\equiv(\theta,\phi)$. To interpret this phenomenon, we study in detail the scattering process of a representative scattering state, as schematically shown in Fig.3(c). When a left-incoming electron with $\xi<1$ for example is incident on the left interface, the transmitted wave consists of two forward evanescent modes corresponding to term $a$ and $b$ in Eq.(\ref{parawave}). Only term $a$ is actually a `propagating' mode which is capable of reaching another interface as long as the length $N$ is finite and $\xi$ is sufficiently close to $1$. When this single mode is incident on the right interface, it will cause a definite transmitted wave as described by Eq.(\ref{paraRleadwave}). The reflected wave by the right interface will then be reflected alternatively by the left and right interfaces. As a result, the amplitude $t^{\sigma}$ of the transmitted wave will be renormalized but with the spin polarization $(\theta,\phi)$ left unaltered. Since the `propagating' mode is an eigenmode of the system, for an incident electron with fixed $\xi$, varying its spin orientation can only change this mode's amplitude $a$ and thus can only change the total transmission amplitude $t^{\sigma}$, keeping the spin polarization unchanged.

Because all the $\bm{k}^{\perp}$ points on the projection loop have the identical $\mathcal{H}^{\parallel}_{\mathrm{eff}}(k_{z})$ with $\xi=1$, electrons incident from them share the same transmitted spin polarization $(\theta,\phi)$. However, when the incident $\bm{k}^{\perp}$ is scanning along the radial direction, $(\theta,\phi)$ varies with $\xi$, which is shown in Fig.\ref{fig4}(a)-(b). The variation of $(\theta,\phi)$ is slight if that of $\bm{k}^{\perp}$ is kept within the annulus. Actually, by considering the scattering state of the $\phi_{+}$ mode being incident on the right interface, one can obtain: $t_{\uparrow\sigma}/t_{\downarrow\sigma}=i(\eta-ie^{ik^{N}_{z}})/(\eta+ie^{ik^{N}_{z}})$, and $(\theta,\phi)$ can be derived analytically: $\theta=\cos^{-1}(\frac{2\eta}{1+\eta^{2}}\sin k^{N}_{z})$, $\phi=\tan^{-1}(\frac{1-\eta^{2}}{2\eta\cos k^{N}_{z}})$, where $\eta=(1+\lambda)/2\gamma^{2}$. Therefore, a nearly fully spin-polarized transmitted current can be expected, which is shown in Fig.\ref{fig4}(c). Here for a NLS system with $N=100$, the current polarization $P=e|\bm{G}^{s}|/\hbar G^{c}$ can reach $99.992\%$. By increasing the length $N$, higher polarization $P$ can be expected since the width of the annulus proportional to $\lambda/$N will become more narrower. Numerical calculations also reveal that for spin-up or spin-down incident electrons, the superposition coefficients of the wave function in NLS Eq.(\ref{parawave}) are proportional to each other and obey the relation $a^{\uparrow}/a^{\downarrow}=c^{\uparrow}/c^{\downarrow}=d^{\uparrow}/d^{\downarrow}$, which is consistent with our interpretation.

The complete spin polarization indicates that the transmission matrix $t$ is singular, since $(t_{\uparrow\uparrow},t_{\downarrow\uparrow})^{T}$ and $(t_{\uparrow\downarrow},t_{\downarrow\downarrow})^{T}$ are proportional to each other because of Eq.(\ref{paraRleadwave}). One can thus change the spin basis of lead R to transform $t$ into a more meaningful form. By rotating the spin axis from $\bm{z}$ to that defined by $(\theta,\phi)$, $t$ becomes $\left(
                                                                       \begin{array}{cc}
                                                                         t^{\uparrow} & t^{\downarrow} \\
                                                                         0 & 0 \\
                                                                       \end{array}
                                                                     \right)
$. According to the relation between the charge and spin current densities for a pair of incident electrons with opposite spin orientations: $\left|\bm{J}^{s}\right|=\sqrt{(\hbar J^{c}/e)^{2}-4\det(t)}$, we have $J^{c}=\frac{e}{\hbar}|\bm{J}_{s}|$.

\begin{figure}[ht]
  \begin{center}
	\includegraphics[width=8.5cm,height=8cm]{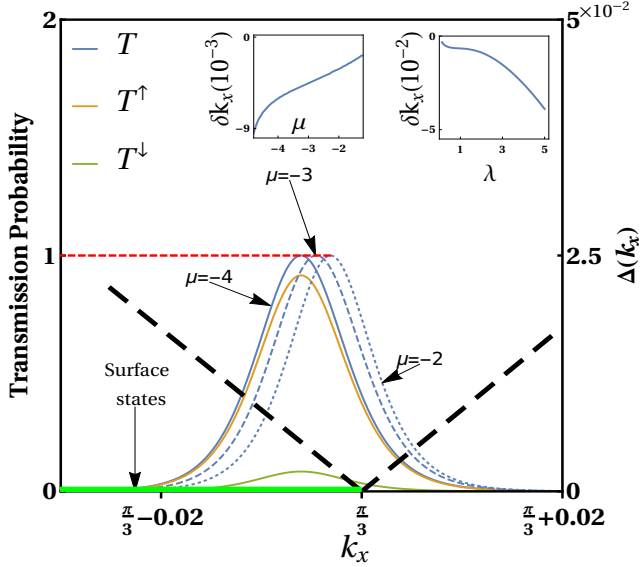}	
  \end{center}
  \vspace{-0.4cm}
  \caption{(Color online) Total transmission probabilities for a pair of incident electrons with opposite spin orientations as their transverse momentum $\bm{k}^{\perp}$ is scanning along $k_{x}$ axis and near point B with $\bm{k}^{\perp}_{B}=(\frac{\pi}{3}, 0)$, as denoted in Fig.\ref{fig3}(b). For different chemical potential $\mu$(or $k_{F}$) of the normal leads, all curves peak at locations near point B. Here $\lambda=0.5$. The insets give the peak location as functions of $\mu$ and $\lambda$ respectively. The thick dashed line is the band gap $\Delta(k_{x})=\frac{\sqrt{3}}{2}|k_{x}-\frac{\pi}{3}|$ and the thick green solid line denotes the drumhead surface states.
  } \label{fig5}
\end{figure}

\subsubsection{Total reflection}
As the transmitted wave Eq.(\ref{paraRleadwave}) for a spin-up incident electron is proportional to that for a spin-down one with identical $\bm{k_{\perp}}$, indicating the two properly superposed transmitted waves could exactly cancel out, the incident electron whose wave function proportional to $(t^{\downarrow},-t^{\uparrow})^{T}$ would be totally reflected. A specific spin orientation $(\theta_{in},\phi_{in})$ for the incident electrons can thus be defined as $(\cos\frac{\theta_{in}}{2} e^{-i\phi_{in}},\sin\frac{\theta_{in}}{2})^{T} \propto (t^{\downarrow},-t^{\uparrow})^{T}$. Because $t^{\uparrow}/t^{\downarrow}=i(\eta+ie^{ik^{N}_{z}})/(\eta-ie^{ik^{N}_{z}})=-t_{\downarrow\sigma}/t_{\uparrow\sigma}=-\tan\theta e^{i\phi}$, $(\theta_{in},\phi_{in})$ is found to obey an interesting relation with the spin-polarization angle $(\theta,\phi)$ of the transmitted wave: $\theta_{in}=\pi-\theta$, $\phi_{in}=\pi-\phi$. By rotating the spin axis of lead L from $\bm{z}$ to that defined by $(\pi-\theta_{in},\pi+\phi_{in})=(\theta,-\phi)$, the transmission matrix $\left(
                                                                       \begin{array}{cc}
                                                                         t^{\uparrow} & t^{\downarrow} \\
                                                                         0 & 0 \\
                                                                       \end{array}
                                                                     \right)
$ can be further transformed into $\left(
                                    \begin{array}{cc}
                                      t^{\nearrow} & 0 \\
                                      0 & 0 \\
                                    \end{array}
                                  \right)$
, where $t^{\nearrow}=\cos\frac{\theta}{2}e^{i\phi}t^{\uparrow}+\sin\frac{\theta}{2}t^{\downarrow}$ is the total transmission amplitude for the incident electron with spin orientation $(\theta,-\phi)$. This effect indicates that the NLS materials could act as spin-valve devices in future spintronics.

\begin{figure}[ht]
  \begin{center}
	\includegraphics[width=8cm,height=6.5cm]{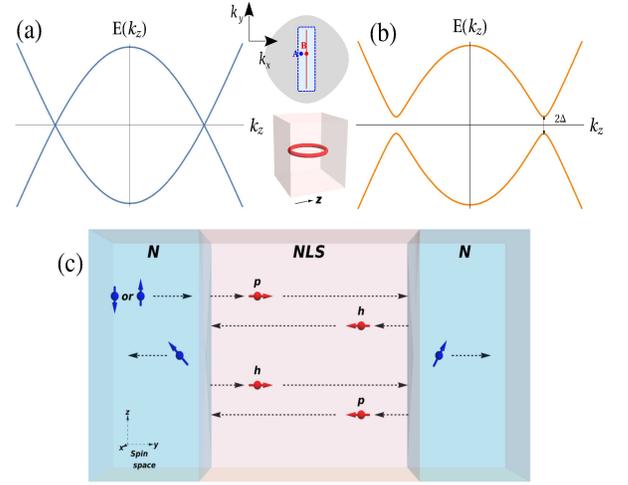}	
  \end{center}
  \vspace{-0.4cm}
  \caption{(Color online)Low-energy dispersion of the effective 1D system of the NLS in the perpendicular case for (a)$k_{x}=0$, and (b)$k_{x}\neq0$. While the former is gapless, the latter has a small gap $\Delta=\lambda |\sin k_{x}|$. The upper inset gives the red projection line of the nodal loop and that of the Fermi sphere of the normal leads. The lower one gives the 3D geometry of the nodal loop($\|\bm{z}$). (c)Schematic of a representative scattering process occurring in the N-NLS-N junction in the perpendicular case. Here the symbol p(h) represents the particle(hole)-like propagating modes.
  } \label{fig6}
\end{figure}
\subsubsection{Resonance of transmission probability and half transmission}
The third novel effect is the resonance of transmission probability. Let us consider a pair of incoming electrons with opposite spin orientations but identical $\bm{k}^{\perp}$. The total transmission probability can be expressed as $\sum_{\sigma',\sigma}|t_{\sigma'\sigma}|^{2}=$Tr$t^{\dagger}t$. Generically, because of Eq.(\ref{leadfunc2}), Tr$t^{\dagger}t=2$-Tr$r^{\dagger}r\leq2$. However, a more strong inequality can be proved: Tr$t^{\dagger}t\leq1$. This is because Tr$t^{\dagger}t$ is an invariant expression which is independent of spin representation. If we denote the spin states with opposite spin orientations $(\theta_{in},\phi_{in})$ and $(\pi-\theta_{in},\pi+\phi_{in})$ as $|\swarrow\rangle$ and $|\nearrow\rangle$ respectively, then Tr$t^{\dagger}t=|t^{\nearrow}|^{2}=|t_{\uparrow\nearrow}|^{2}+|t_{\downarrow\nearrow}|^{2}=1-|r_{\uparrow\nearrow}|^{2}-|r_{\downarrow\nearrow}|^{2}\leq1$ since $t_{\uparrow\swarrow}=t_{\downarrow\swarrow}=0$. When the incident $\bm{k}^{\perp}$ is scanning along $k_{x}$ axis near $k_{x}=\frac{\pi}{3}$, the transmission probability Tr$t^{\dagger}t$ at different $k_{F}$ are shown in Fig.\ref{fig5}. Resonance of Tr$t^{\dagger}t$ occurs for each curve. Peak position weakly depends on $k_{F}$ but remarkably half transmission Tr$t^{\dagger}t=1$ occurs, i.e., all peak values take exactly value of 1. The fact that the peak position is near and within the projection of the nodal loop is an evidence of participation of the drumhead surface states in the transport process. This can be understood as follows. On one hand, the metallic surface states are expected to give significant contributions to the transport. On the other hand, as $\bm{k}^{\perp}$ is moving away from the projection loop, the gap increases linearly, leading to the suppression of their contribution. As a result of the combination of the two effects, the peak near and within the the projection loop can be expected, as exhibited in Fig.\ref{fig5}.

\subsection{Nodal loop perpendicular to the interfaces}
Secondly, we discuss the transport properties of the NLS in the N-NLS-N junction when the nodal loop is perpendicular to the interfaces. In this situation, the nodal loop of the NLS described by $\cos k_{y}+\cos k_{z}=1.5$ is located at $k_{x}=0$ plane, which can be obtained by rotating the model system of (\ref{paraNLS}) about $\bm{y}$ axis by $90^{\circ}$, or just exchanging $k_{z}$ with $k_{x}$ in the model. Thus the effective 1D Hamiltonian can be given as
\begin{equation}
  \mathcal{H}^{\perp}_{\mathrm{eff}}(k_{z}) = (\xi - \cos k_{z})\sigma_{x} + \lambda \sin k_{x}\sigma_{z},
  \label{perpH}
\end{equation}
where $\xi=2.5-\cos k_{x}-\cos k_{y}$. Its energy spectrum is $E^{\perp}(k_{z})=\pm\sqrt{(\xi-\cos k_{z})^{2}+\lambda^{2}\sin^{2} k_{x}}$. The gapless(gapful) low-energy dispersion for $k_{x}=0$($k_{x}\neq0$) is shown in Fig.\ref{fig6}(a)((b)).

\begin{figure}[ht]
  \begin{center}
	\includegraphics[width=8.5cm,height=9cm]{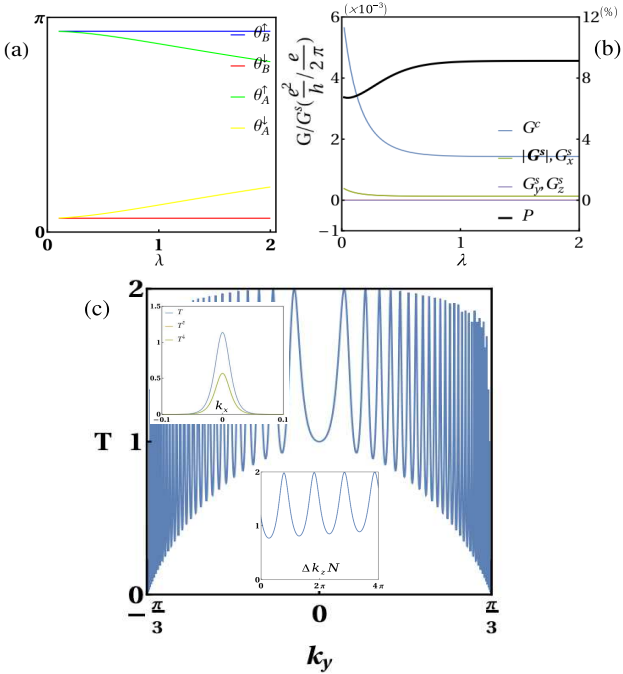}	
  \end{center}
  \vspace{-0.4cm}
  \caption{(Color online)(a)Spin-polarization angle $\theta$ of the transmitted wave as a function of $\lambda$ for two representative incident points shown in the upper inset of Fig.\ref{fig6}. (b)Transmitted charge and spin conductances, as well as current polarization as functions of parameter $\lambda$. (c)Total transmission probability Tr$t^{\dagger}t$ for a pair of incident electrons with identical $\bm{k}^{\perp}$ but opposite spin orientations when $\bm{k}^{\perp}$ is scanning along the projection red line. The lower inset is part of the transmission probability but as a function of the variation of $k^{N}_{z}N$. The upper inset gives that as a function of $k_{_{x}}$ near $k_{x}=0$, when $\bm{k}^{\perp}$ is scanning along $k_{x}$ axis. Here $\lambda=0.5$.
  } \label{fig7}
\end{figure}
\subsubsection{Partially spin-polarized current}
Different from the parallel case, the projection of the nodal loop in this situation is a line segment, as shown in the inset of Fig.\ref{fig6}(a). Similar to the previous discussion, the main contributions to the transport process come from the incident electrons with their $\bm{k}^{\perp}$ near the projection line. Therefore, only regime with $k_{x}\sim0$ is needed to be considered. On the other hand, although $\xi$ shares the same expression to the parallel case, here it can vary within a finite region: $0.5\leq\xi\leq1$. Thus there exist 4 eigensolutions for 1D $\mathcal{H}^{\perp}_{\mathrm{eff}}(k_{z})$: two forward(backward)-propagating modes including particle-like one $k_{z}=k+i\chi(-k-i\chi)$ and hole-like one $k_{z}=-k+i\chi(k-i\chi)$, where $k=\cosh^{-1}\xi$ and $\chi=\lambda k_{x}/\sqrt{1-\xi^{2}}$. See Appendix A for detail. The wave function of the quasi-1D NLS in the junction can be given by,
\begin{equation}
  \begin{split}
	\Psi^{S}(j)&= (a e^{- i k j} + b e^{ i k j }) e^{-\chi j}\phi_{+}+(c e^{ -i k j} + d e^{ i k j})e^{\chi j}\phi_{-},
  \end{split}
  \label{perpW}
\end{equation}
where $a,b,c,d$ are the superposition coefficients. A representative scattering process is schematically shown in Fig.\ref{fig6}(c). Quite different from the parallel case within the NLS region, the scattering state has more than one forward-propagating mode, leading partially spin-polarized transmitted wave. This also indicates that for each relevant scattering state, unlike the parallel case, the transmission matrix $t$ is generically nonsingular. The spin-orientation angle $\theta$ for two representative incident points are exhibited in Fig.\ref{fig7}(a), in which a quite large splitting of the angles between spin-up and spin-down incident electrons is found. The polarization $P$ of this partially spin-polarized transmitted current as well as its spin and charge conductances are shown in Fig.\ref{fig7}(b), giving $P\sim10\%$, much smaller than that of the parallel case.

\subsubsection{Resonance of transmission probability and perfect transmission}
The total transmission Tr$t^{\dagger}t$ of a pair of incident electrons with opposite spin orientations, however, shows periodic resonance behavior, as can be seen in Fig.\ref{fig7}(c). This is interpreted as that the scanning of $\bm{k}^{\perp}$ along the projection line will lead to the variation of the wave vector $k_{z}$ of the quasi-1D NLS, which will then induce a transmission resonance peak whenever the increase of $k_{z}N$ becomes a multiple of $\pi$, as exhibited in the lower inset. When $k_{y}$ is near $0$, perfect transmission Tr$t^{\dagger}t=2$ can actually be realized, which is distinct to the half transmission behavior in the parallel case. The upper inset of Fig.\ref{fig7}(c) also gives Tr$t^{\dagger}t$ as a function of $k_{x}$, when the incident $\bm{k}^{\perp}$ is scanning along $k_{x}$ axis. Note that similar behavior was also observed in the NLS state in the hyperhoneycomb lattice\cite{guanBarrier2018}. Different from the parallel case, Tr$t^{\dagger}t$ peaks at $k_{x}=0$ which is on the projection line and the peak value is between $1$ and $2$. These differences can be attributed to the absence of the drumhead surface states in this situation.

\begin{figure}[ht]
  \begin{center}
	\includegraphics[width=9cm,height=6cm]{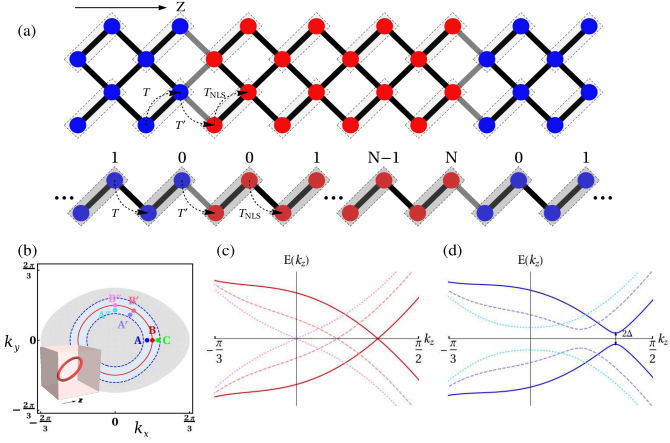}	
  \end{center}
  \vspace{-0.4cm}
  \caption{(Color online) (a)Microscopic structure of the N-NLS-N junction, when the nodal loop is intersecting the interfaces at $45^{\circ}$. The effective quasi-1D lattice model is also given below. (b)The projection of the nodal loop on $k_{z}=0$ plane, denoted by the solid line. The elliptical grey region is the projection of the Fermi sphere of the normal leads. The inset gives the 3D geometry of the nodal loop(intersecting $\bm{z}$ at $45^{\circ}$). Low-energy spectra of (c)gapless and (d)gapful quasi-1D NLS for the three representative incoming points $BB'B''$ and $AA'A''$ denoted in (b) respectively.
  } \label{fig8}
\end{figure}

\begin{figure}[ht]
  \begin{center}
	\includegraphics[width=8.5cm,height=8cm]{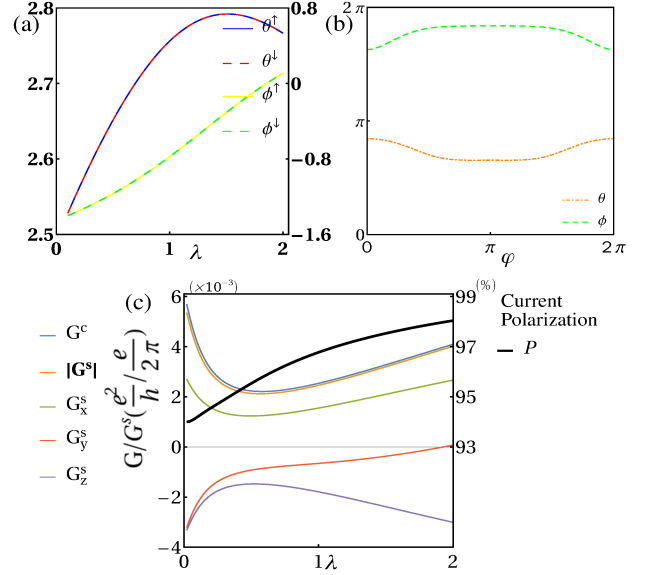}	
  \end{center}
  \vspace{-0.4cm}
  \caption{(Color online) (a)Spin-polarization angle $(\theta,\phi)$ of the transmitted wave for the incident point B denoted in Fig.\ref{fig8}(b) as functions of parameter $\lambda$. Here the nodal loop is intersecting the interfaces at $45^{\circ}$. The spin polarization is also found to be independent of incident electron's spin orientation, so $\theta^{\uparrow}=\theta^{\downarrow},\phi^{\uparrow}=\phi^{\downarrow}$. (b)Spin-polarization angle $(\theta,\phi)$ when the incident $\bm{k}^{\perp}$ is scanning along the projection loop. Here $\varphi=0$ corresponds to point B and $\lambda=0.5$. (c)Transmitted charge and spin conductances, as well as current polarization as functions of parameter $\lambda$.
  }
 \label{fig9}
\end{figure}

\subsection{Nodal loop intersecting the interfaces at \texorpdfstring{$45^{\circ}$}{Lg}}
We now turn to the case where the nodal loop is intersecting the interfaces at $45^{\circ}$ in the N-NLS-N junction. This situation can be realized by rotating both the NLS model system (24) and the normal leads about $\bm{y}$ axis by $45^{\circ}$, as schematically shown in Fig.\ref{fig8}(a). Thus each unit cell of the corresponding quasi-1D system contains two atoms, leading to two bands for each spin index in normal leads, which can be written by

\begin{equation}
  E^{45^{\circ}}(k^{N}_{z})=-2 \cos k_{y} -\mu \pm 4 \cos\frac{k_{x}}{2}\cos\frac{k^{N}_{z}}{2}.
\end{equation}
In general, a four-band system often has four forward modes. However, two forward modes are absent in this case and only two ones are left. Similar conclusion can be made to the NLS in this situation: There only exist two forward-propagating modes, of which when $\bm{k}^{\perp}$ is near the projection loop, one is particle-like(or hole-like) and nearly propagating while the other is evanescent. Since there is only a single forward-propagating mode, exactly similar mechanism to the parallel case will lead to the conclusion that for any scattering state, the transmitted wave will be fully spin polarized. But unlike the parallel case, the reflected backward-propagating mode is replaced here by a hole-like(or particle-like) one. For a representative scattering state, we demonstrate the complete spin polarization of the transmitted wave in Fig.\ref{fig9}(a). Similarly, for each scattering state, there is a special incident spin orientation $(\theta_{in}, \phi_{in})$ corresponding to the total reflection, and the resonance of transmission probability Tr$t^{\dagger}t$ and half transmission also occurs. Furthermore, the incident points on the projection loop with different $\bm{k}^{\perp}$ are described by different effective 1D Hamiltonian, resulting in different spin orientations of the transmitted waves, as shown in Fig.\ref{fig9}(b). This then leads to the nearly fully spin-polarized transmitted wave, as the current polarization $P$ can be varying between $94\%$ and $98\%$ in the parameter space studied, as can seen in Fig.\ref{fig9}(c).

\subsection{Spin current density and spin torque in the NLS}
Now we analyze the nonconserved spin current density and spin torque in the NLS of the junction introduced in Sec.\ref{sect2}. We take the parallel case to demonstrate the novel spin transport. According to Eq.(19), and by reexpressing the site-independent $\epsilon_{j}$ and $T_{j,j+1}$ as $\epsilon_{j}=\bm{h}\cdot\bm{\sigma}$, $T_{j,j+1}=\bm{t}_{s}\cdot\bm{\sigma}=T^{*}_{j+1,j}$, with $\bm{h}=(\xi,0,0)$ and $\bm{t}_{s}=(-1/2,0,-i\lambda/2)$, the spin torque $\bm{g}(j)$ can be rewritten as,
\begin{equation}
  \begin{split}
	\bm{g}(j) &=-\operatorname{Re}\{\Psi^{\dagger}(j)\bm{\sigma}\times[\bm{h}\Psi(j) \\
	&+\bm{t}_{s}\Psi(j+1)+\bm{t}^{*}_{s}\Psi(j-1)]\}.	
  \end{split}
  \label{source1}
\end{equation}
Here $\Psi(-1)=\Psi(N+1)\equiv0$. For an arbitrary scattering state, it can be exactly proved that $g_{x}(j)=0$, $g_{y}(j)=0$, where the latter one is obeyed except for $j=0$ or $N$. The only nonzero term is $g_{z}(j)$, which takes relatively larger value only near the interfaces while takes the form $\bm{g}_{z}(j)=2(\xi-1)(|a|^{2}e^{-2\chi_{1}j}-|c|^{2}e^{-2\chi_{1}(N-j)})$ when away from the interfaces, being rather small, as shown in Fig.10. The spin current $\bm{J}^{s}(j)\equiv\bm{J}^{s}_{j+1\leftarrow j}=\bm{J}^{s}_{j+1,j}$ can be derived as $\bm{J}^{s}(j)=\operatorname{Im}\{\bm{t}_{s}\Psi^{\dagger}(j+1)\Psi(j)\}$, which means $J^{s}_{y}(j)=0$. Actually, we even have $J^{s}_{x}(j)=0$ and $J^{s}_{z}(j)$ can also be given analytically by $J^{s}_{z}(j)=-\lambda(|a|^{2}e^{-\chi_{1}(2j+1)}+|c|^{2}e^{-\chi_{1}(2N-2j-1)})/2$ when away from the interfaces. For a pair of incident electrons with identical $\bm{k}^{\perp}$ but opposite spin orientations, their contributions to the spin currents in lead $R$ is $\bm{J}^{s}_{R}=|t^{\nearrow}|^{2}(\sin\theta\cos\phi,\sin\theta\sin\phi,\cos\theta)$. Notice that electrons with the same $\bm{k}^{\perp}$ but incident from lead R will also be transmitted to lead L with their spin nearly fully polarized at $(\theta',\phi')$. Because of the symmetry of the transport system, if we make a spin rotation about $\bm{x}$ by $180^{\circ}$($\sigma_{y}\rightarrow-\sigma_{y}$, $\sigma_{z}\rightarrow-\sigma_{z}$ ), the right incident scattering states can be equivalent to the left ones. This implies that the spin-polarization angle $(\theta',\phi')$ satisfies: $\theta'=\pi-\theta$, $\phi'=-\phi$. Thus for the pair of electrons, their contributions to the spin current in lead L is $\bm{J}^{s}_{L}=|t^{\nearrow}|^{2}(\sin\theta'\cos\phi',\sin\theta'\sin\phi',\cos\theta')=|t^{\nearrow}|^{2}(\sin\theta\cos\phi,-\sin\theta\sin\phi,-\cos\theta)$.
Although $|\bm{J}^{s}_{R}|=|\bm{J}^{s}_{L}|$, $\bm{J}^{s}_{R}\neq\bm{J}^{s}_{L}$, which means that the spin torque existing in the NLS plays a role of transforming the spin current in lead L to that in lead R. If we denote the increase of the spin current due to the spin torque by the two interfaces as $\delta\bm{J}^{s}_{L(R)}$, we have $\delta J^{s}_{L,x}=-\delta J^{s}_{R,x}$, and $\delta J^{s}_{L,y}=\delta J^{s}_{R,y}$. For $\delta J^{s}_{z}$, the difference $\delta J^{s}_{R,z}-\delta J^{s}_{L,z}$ actually equals the sum of the small spin torque $g_{z}(j)$ over the whole NLS region. Finally, for the spin conductance, the above argument also leads to: $(G^{s}_{L,x}, G^{s}_{L,x},G^{s}_{L,z})=(G^{s}_{R,x}, -G^{s}_{R,y},-G^{s}_{R,z})$.

\begin{figure}[ht]
  \begin{center}
	\includegraphics[width=8.5cm,height=3.8cm]{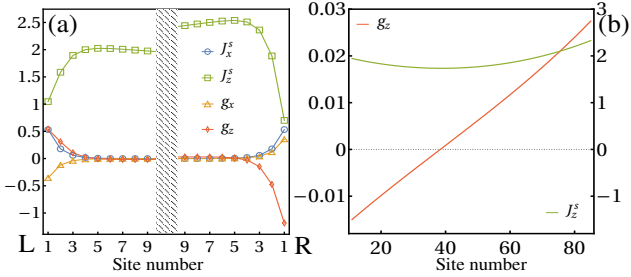}	
  \end{center}
  \vspace{-0.4cm}
  \caption{(Color online) The spatial distributions of nonzero spin current density and spin torque in the NLS of the junction for the parallel case, where the scattering state is for a pair of incident electrons from lead L with identical $\bm{k}^{\perp}$ but opposite spin orientations. (a)Near the interfaces. (b)Away from the interfaces. Here $\lambda=0.5$ and $\bm{k}^{\perp}=(\pi/3-0.005,0)$.
  } \label{fig10}
\end{figure}
\begin{table*}[ht]
  \caption{Classification of wave vector $k_{z}$ in the quasi-1D NLS in the parallel case. Here $f_{\pm}(\xi,\lambda)\equiv\frac{\xi\pm\sqrt{\xi^{2}+\lambda^{2}-1}}{1-\lambda}$ and $\lambda>0$.}
\label{tab:table1}
  \begin{tabular}{lccccc|ccc}
\hline \hline
\multicolumn{1}{c}{\multirow{3}{*}{}} & \multicolumn{8}{c}{Nodal loop $\parallel$ the interfaces}                               \\ \hline
\multicolumn{1}{c}{}                  & \multicolumn{5}{c|}{$0<\lambda< 1$} & \multicolumn{3}{c}{$\lambda> 1$} \\
\multicolumn{1}{c}{}                  & $\sqrt{1-\lambda^{2}}<\xi$ &  $-\sqrt{1-\lambda^{2}}>\xi$  & $\left|\xi\right|<\sqrt{1-\lambda^{2}}$  & $\xi-1\sim0$ & $\left|\xi\right|=\sqrt{1-\lambda^{2}}$ &  $\xi>0$   &  $\xi-1\sim0$    &  $\xi<0$     \\  \hline
   $k_{1}$ &  $0$    &  $\pi$    & $\cos^{-1}\frac{r}{\sqrt{1-\lambda^{2}}}$  & $0$  & $0$  &  $0$  &  $0$   &  $\pi$     \\

   $\chi_{1}$ &  $\ln f_{-}(\xi,\lambda)$ & $\ln f_{-}(-\xi,\lambda)$  &  $\tanh^{-1} \lambda$   & $\frac{1-\xi}{\lambda}$ & $\frac{1}{2}\ln \frac{1+\lambda}{1-\lambda}$ &  $\makecell{\ln f_{-}(\xi,\lambda)}$   & $\frac{1-\xi}{\lambda}$   &  $\ln \{-f_{+}(\xi,\lambda)\}$     \\

$k_{2}$ & $0$  &  $\pi$    &  $-\cos^{-1}\frac{r}{\sqrt{1-\lambda^{2}}}$   & $0$ & $0$ &   $\pi$     &   $\pi$     &   $0$  \\

$\chi_{2}$ &  $\ln f_{+}(\xi,\lambda)$    &  $\ln f_{+}(-\xi,\lambda)$   &  $\tanh^{-1} \lambda$   &  $\ln\frac{1+\lambda}{1-\lambda}$ & $\frac{1}{2}\ln \frac{1+\lambda}{1-\lambda}$  & $\makecell{\ln \{-f_{+}(\xi,\lambda))\}}$    &  $\ln\frac{1+\lambda}{1-\lambda}$      &  $\ln f_{-}(\xi,\lambda)$   \\ \hline \hline
\end{tabular}
\end{table*}
\subsection{Generalization to the multi-band NLSs and discussion}

\begin{figure}[ht]
  \begin{center}
	\includegraphics[width=8cm,height=6cm]{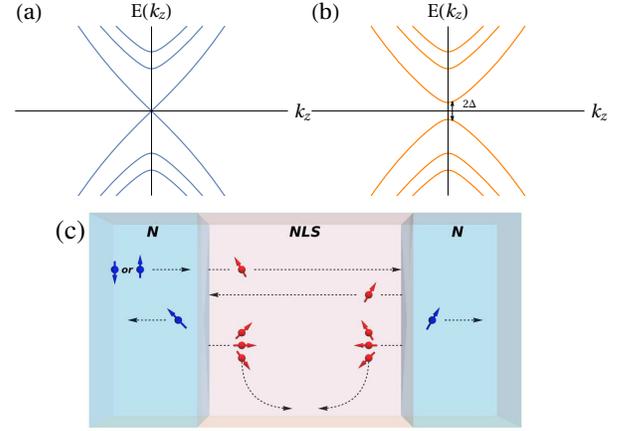}	
  \end{center}
  \vspace{-0.4cm}
  \caption{(Color online) Schematic low-energy dispersion of the effective 1D multi-band NLS when the nodal loop is parallel to the interfaces, with $\bm{k}^{\perp}$ (a) on or (b) near the projection of the nodal loop. (c)Schematic of the scattering mechanism occurring in the N-NLS-N junction in this situation. As before, the colored arrows denote spin orientations while the black straight(curved) arrows denotes the propagating(evanescent) modes.
  } \label{fig11}
\end{figure}

All our conclusions above on the NLS including the complete spin polarization, total reflection and half transmission, are based on the minimal model Eq.(\ref{paraNLS}). We now argue that most of these phenomena are model independent and should be the general features of the Weyl NLSs possessing one single doubly degenerate nodal loop. To demonstrate this point, we consider a 2$n$-band($n\geq2$) Weyl NLS with the nodal loop parallel to the interfaces. The doubly degenerate nodal loop can be seen as the intersection between the two lowest-energy bands. The energy spectrum of the quasi-1D NLS in the junction also has two low-energy bands, as schematically shown in Fig.\ref{fig11}(a)-(b). Thus when the incident $\bm{k}^{\perp}$ is on the projection loop, among the $n$ forward modes, there is a propagating one with $k_{z}=0$ whereas the others are evanescent ones. Then when $\bm{k}^{\perp}$ is close to the projection loop, the propagating mode becomes nearly propagating with $k_{z}$ being a quite small imaginary number. As a result, only the nearly propagating one can reach the right interface(Fig.\ref{fig11}(c)). The other $n-1$ evanescent forward modes do not play an important role in the transport. As long as each incident $\bm{k}^{\perp}$ on the projection loop shares the identical or similar qusi-1D effective Hamiltonian, the existence of this one single forward-propagating mode will lead to the conclusions mentioned above according to similar argument. The doubly degenerate Weyl nodal loop is crucial here, since a four-fold degenerate Dirac NLS for example would result in two forward-propagating modes, which would then lead to trivial conclusions.

The doubly degenerate nodal fermions can be expected to be realized in SO coupled or ferromagnetic materials such as PbTaSe$_{2}$\cite{bianTopological2016} and Co$_{2}$MnGa\cite{belopolskiDiscovery2019}. But we note that our conclusions cannot be applied directly to these two Weyl NLS materials, because both of them have multiple nodal loops, resulting in elimination of the spin polarization of the transmitted current. We also note that in Weyl NLS ferromagnetic materials Li$_{3}$(FeO$_{3}$)$_{2}$\cite{chenWeylloop2019} and Fe$_{3}$GeTe$_{2}$\cite{kimLarge2018}, the spin degree of freedom is fully quenched by the large ferromagnetic polarization and thus these two half -metallic materials can be viewed as spinless ones, indicating that they still cannot be served as the candidate materials.

In all the above discussions, the Fermi level is fixed at the nodal line of the NLS, i.e., $E=0$. A slight deviation of $E$ from $0$ in the actual situation would lead to the replacement of the nodal line by a 2D torus Fermi surface. We now argue that this does not change any of the above main results. The projection of the 2D torus on the interfaces is an annulus. When the transverse momentum $\bm{k_{\perp}}$ of an incident electron is within the annulus, besides many evanescent modes, there exist two propagating ones with real $k_{z}$ solutions given by $k_{z}=\pm\cos^{-1}(\frac{\xi-\sqrt{\xi^{2}\lambda^{2}+(\lambda^{2}-1)(\lambda^{2}-E^{2})}}{1-\lambda^{2}})$, among which, only one is forward-propagating, leading to similar main results mentioned above.

\section{Summary}
\label{sect4}
We have introduced a wave-function method in lattice form to study transport properties of Weyl NLSs. This method gives directly the wave function of the scattering region, based upon which, we have further derived the charge and spin conservation laws and currents. Our study on a junction made up of a Weyl NLS and normal metals indicates that the Weyl NLSs possessing a single nodal loop parallel to the junction interfaces have novel spin transport properties: Incident electrons with a special spin orientation would be totally reflected. The surface-state involved half transmission occurs as the transmission resonance. The transmitted charge current is nearly fully spin-polarized. These phenomena can be attributed to the existence of only one forward-propagating mode in the NLS of the junction. This picture is found to be model independent and has been generalized to the case of multi-band Weyl NLSs. All these features are expected to be verified by future transport experiments and would be also helpful in detecting new Wely NLS materials.

\begin{acknowledgments}
Y.Z. thanks Qingli Zhu for useful discussions. This work is supported by NSFC Project No.11874202 and 973 Project No.2015CB921202.
\end{acknowledgments}

\begin{appendix}
\section{Eigensolutions of the quasi-1D NLS}

\begin{table}[ht]
  \caption{Classification of wave vector $k_{z}$ in the quasi-1D NLS in the perpendicular case. Only the most interesting regime for which $k_{x}\sim0$ is given here.}
\label{tab:table2}
\begin{ruledtabular}
\begin{tabular}{ccccc}
\multirow{3}{*}{} & \multicolumn{3}{c}{Nodal loop $\perp$ the interfaces}       \\  \colrule
& \multicolumn{3}{c}{$k_{x}\sim 0$}    \\
  &  $\left|\xi \right|\leq 1$     & $\xi>1$    & $\xi<-1$         \\ \colrule
  $k$ & $\sim\frac{\pi}{2}-\sin^{-1}\xi$      &  $\sim \frac{\lambda k_{x}}{\sqrt{\xi^{2}-1}}$             & $\sim \pi-\frac{\lambda k_{x}}{\sqrt{\xi^{2}-1}}$                              \\
$\chi$ & $\sim\frac{\lambda k_{x}}{\sqrt{1-\xi^{2}}}$            &  $\sim \cosh^{-1}\xi$   & $\sim \cosh^{-1}\left|\xi\right|$           \\
\end{tabular}
\end{ruledtabular}
\end{table}

\label{appe}
\label{App:para}
In this appendix we give in detail the eigenmodes of the quasi-1D NLS in the N-NLS-N junction. We first consider the parallel case where the nodal loop is parallel to the interfaces, discussed in Sec.\ref{sect3}A. All solutions of $k_{z}$ together with their corresponding eigenmodes can be obtained by solving the eigenequation of Eq.(\ref{paraEH}) for $E=0$. It is found that there are always $4$ solutions of $k_{z}$ in total and the propagating modes($k_{z}=0$) exist only for $\xi=1$. Thus $k_{z}$ of a generic solution has to be complex and can be expressed as $k+i\chi$. Here $k$ and $\chi$ are real variables obeying,
\begin{equation}
  \begin{split}
	\sin k(\sinh \chi\pm\lambda \cosh \chi)=0, \\
	\cos k(\cosh \chi\pm\lambda\sinh \chi)=\xi.
\end{split}
\end{equation}
The symbol `$\pm$' means that the above two equations take `$+$' or `$-$' sign simultaneously. Without loss of generality, we assume $\lambda>0$. If $k_{z}$ is a solution for eigenmode $\phi$, $k^{*}_{z}$ is also one for eigenmode $\phi^{*}$, leading to the $4$ solutions of $k_{z}$: $k_{1}+i\chi_{1}$, $k_{2}+i\chi_{2}$ corresponds to $\phi_{+}=\frac{1}{\sqrt{2}}(1,i)^{T}$, and $k_{1}-i\chi_{1}$, $k_{2}-i\chi_{2}$ corresponds to $\phi_{-}=\frac{1}{\sqrt{2}}(1,-i)^{T}$ with $|\chi_{1}|\leq|\chi_{2}|$. All kinds of solutions for $k_{1}$, $k_{2}$, $\chi_{1}$ and $\chi_{2}$ at different situations are summarized in the Tab.\ref{tab:table1}. The wave function in the NLS can be generally expanded as in Eq.(\ref{parawave}) in the main text. For any relevant scattering state described in Eq.(\ref{leadfunc1})-(\ref{leadfunc2}), these coefficients together with $r_{mn}$ and $t_{mn}$ can be determined by solving the following $4$ Schr\"odinger equations for the unit cells near the interfaces,
\begin{equation}
  \begin{split}
	T\Psi^{L}(1)+H_{L}(0)\Psi^{L}(0)+T'^{\dagger}_{a}\Psi^{S}(0) & = E\Psi^{L}(0),\\
    T'_{a}\Psi^{L}(0)+H_{S}(0)\Psi^{S}(0)+T_{S}\Psi^{S}(1) & = E\Psi^{S}(0),\\		
    T_{S}^{\dagger}\Psi^{S}(N-1)+H_{S}(N)\Psi^{S}(N)+T'_{b}\Psi^{R}(0) & = E\Psi^{S}(N),\\
	T'^{\dagger}_{b}\Psi^{S}(N)+H_{R}(0)\Psi^{R}(0)+T\Psi^{R}(1) & = E\Psi^{R}(0),\\
  \end{split}
  \label{coef}
\end{equation}
where $T_{S}$ is the NN hopping matrix in the NLS and $T'$ is that connecting the normal leads to the NLS. The number of layers of the NLS is $N+1$. If all the wave functions could be expressed analytically, by extending the range of $j$ in $\Psi^{L}$, $\Psi^{R}$ from $j\geq0$ to $j\geq-1$, and $j$ in $\Psi^{S}$ from $N\geq j\geq0$ to $N+1\geq j\geq-1$, the above equations can actually be greatly simplified,
\begin{equation}
  \begin{split}
	T^{\dagger}\Psi^{L}(-1)&=T'^{\dagger}_{a}\Psi^{S}(0),T'_{a}\Psi^{L}(0)=T^{\dagger}_{S}\Psi^{S}(-1),\\		
    T^{\dagger}\Psi^{R}(-1)&=T'^{\dagger}_{b}\Psi^{S}(N),T'_{b}\Psi^{R}(0)=T_{S}\Psi^{S}(N+1).\\
  \end{split}
\end{equation}

Next, we consider the perpendicular case where the nodal loop is perpendicular to the interfaces, discussed in Sec.\ref{sect3}B. In an exactly similar way, by solving the eigenequation of Eq.(\ref{perpH}) for $E=0$, the 4 solutions of $k_{z}$ can be obtained: $k_{z}=\pm k \pm i \chi$, where $\pm k+i \chi(\pm k-i \chi)$ corresponds to eigenmode $\phi_{+}(\phi_{-})$. In this case, $k_{x}$ is a good quantum number acting as a varying parameter. It is found that $\phi_{\pm}=\frac{1}{\sqrt{2}}(1,\pm i)^{T}$, when $k_{x}\neq 0$, and $\phi_{\pm}=\frac{1}{\sqrt{2}}(0,\pm1)^{T}$ otherwise. $k$ and $\chi$ are found to obey
\begin{equation}
  \begin{split}
	\sin k\sinh \chi &= \lambda\sin k_{x}, \\
	\cos k\cosh \chi &=\xi.
\end{split}
\end{equation}
All solutions for $k$ and $\chi$ are summarized in the Tab.\ref{tab:table2}. Then the wave function in this case can be given by Eq.(\ref{perpW}) in the main text, where the coefficients $a,b,c,d$ can be determined similarly according to Eq.(\ref{coef}).

\end{appendix}

\bibliographystyle{apsrev4-1}
\bibliography{ref}

\end{document}